\documentclass[preprint,showpacs,superscriptaddress,amsmath,amssymb,prl]{revtex4}

\usepackage{graphicx}
\usepackage{dcolumn}
\usepackage{bm}

\begin{document}

\title{Virtual turning points and bifurcation of Stokes curves
for higher order ordinary differential equations}

\author{Takashi AOKI}
\affiliation{Department of Mathematics, Kinki University, 
Higashi-Osaka, 577-8502 Japan}
\author{Takahiro KAWAI}
\affiliation{Research Institute for Mathematical Sciences, 
Kyoto University, Kyoto, 606-8502 Japan}
\author{Shunsuke SASAKI}
\affiliation{Research Institute for Mathematical Sciences, 
Kyoto University, Kyoto, 606-8502 Japan}
\author{Akira SHUDO}
\affiliation{Department of Physics, Tokyo Metropolitan University, 
Hachioji, Tokyo 192-0397 Japan}
\author{Yoshitsugu TAKEI}
\affiliation{Research Institute for Mathematical Sciences, 
Kyoto University, Kyoto, 606-8502 Japan}

\date{\today}

\begin{abstract}
For a higher order linear ordinary differential operator $P$, 
its Stokes curve bifurcates in general when it hits another
turning point of $P$.
This phenomenon is most neatly understandable by taking 
into account Stokes curves emanating from virtual turning
points, together with those from ordinary turning points. 
This understanding of the bifurcation of a Stokes curve plays
an important role in resolving a paradox recently found in the 
Noumi-Yamada system, a system of linear differential equations 
associated with the fourth Painlev\'e equation.
\end{abstract}

\maketitle

As is pointed out by Silverstone \cite{1},
notorious ambiguities in the connection problems in WKB theory
are resolved if we make use of the Borel resummation method;
in a word, we have to first specify the region (the so-called
Stokes region) where the Borel sum of a WKB solution is well-defined.
The importance of the Borel resummation in WKB analysis is also shown
from several viewpoints by Bender-Wu, Voros, Zinn-Justin and others 
\cite{2}. In the description of the Stokes region for a second order 
linear ordinary differential operator 
$P=P(x,\eta^{-1}d/dx)=P(x,\eta^{-1}\xi)$ with a large parameter $\eta$
we need to consider only
Stokes curves emanating from turning points, that is, the
union of an integral curve of the direction field
\begin{equation}\label{eq:2}
{\rm Im}(\xi_j(x)-\xi_k(x))dx=0
\end{equation}
that emanates from a point {\it a} satisfying
\begin{equation}\label{eq:3}
\xi_j(a)=\xi_k(a)
\end{equation}
where $\xi_j(x)$ and $\xi_k(x)$ are characteristic roots of the
operator $P$.
For higher order operators, however, the totality of Stokes curves 
emanating from turning points (i.e., points satisfying \eqref{eq:3})
does not suffice to describe the Stokes region as Berk et al.
\cite{3} first pointed out; we need a ``new Stokes curve'' that
does not emanate from a turning point in the complete description 
of the Stokes region. Later three of the ${\rm authors}$
\cite{4} noticed that a new Stokes curve is an ordinary Stokes
curve (i.e., an integral curve of \eqref{eq:2}) that emanates from a 
virtual turning point, which is defined through the Borel transform
$P_B$ of the operator $P$, i.e., $P_B=P(x,\partial_{x}/\partial_{y})$.
For example, let us consider the following third order operator 
$P$ \cite{3}:
\begin{equation}\label{eq:4}
P=\eta^{-3}\frac{d^3}{dx^3}+3\eta^{-1}\frac{d}{dx}+2ix
\qquad (\eta \gg 1)
\end{equation}

%%%%%%%%%%%%%%%%%%%%%%%%%%%%%%%%%%%%%%%%%%%%%%%%%%%%%%%%%%%%%%%%%%%%%%%%%%%%%%%
\begin{figure}[h]
\begin{center}
\includegraphics[width=0.25\linewidth]{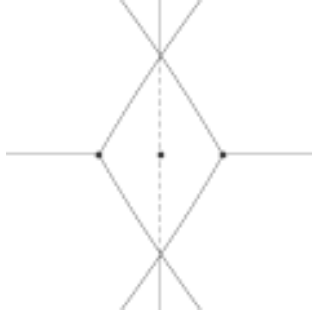}
\end{center}
\caption{Stokes geometry for the operator given by \eqref{eq:4}.
}
\label{bnr0}
\end{figure}
%%%%%%%%%%%%%%%%%%%%%%%%%%%%%%%%%%%%%%%%%%%%%%%%%%%%%%%%%%%%%%%%%%%%%%%%%%%%%%%

With an appropriate labeling of the characteristic root $\xi_j(x)$
of the equation
\begin{equation}\label{eq:5}
\xi^3+3\xi+2ix =0
\end{equation}
we find Fig.~1 of Stokes geometry of $P$;
in Fig. 1, the point $x=-1$ (resp., $+1$) is an ordinary turning
point determined by $\xi_1(x)=\xi_2(x)$ (resp., $\xi_2(x)=\xi_3(x)$),
the point $x=0$ is the virtual turning point detected by Ref. \cite{4};
it satisfies 
\begin{equation}\label{eq:6}
\int^x_{-1}\xi_1(x)dx=\int^{1}_{-1}\xi_2(x)dx+\int^x_1\xi_3(x)dx.
\end{equation}
The Stokes curve defined by the direction field
${\rm Im}(\xi_1(x)-\xi_3(x))dx=0$ that emanates from $x=0$
coincides with the new Stokes curve detected in Ref. \cite{3}.
No Stokes phenomena occur on the dotted portion
of the curve in Fig.~1 (\cite{3}, \cite{4}).
See \cite{4b} for an application to the Landau-Zener type
level crossing problem of the notion of 
virtual turning points and Stokes curves emanating from them.

%%%%%%%%%%%%%%%%%%%%%%%%%%%%%%%%%%%%%%%%%%%%%%%%%%%%%%%%%%%%%%%%%%%%%%%%%%%%%%%
\begin{figure}[h]
\begin{center}
\includegraphics[width=1.0\linewidth]{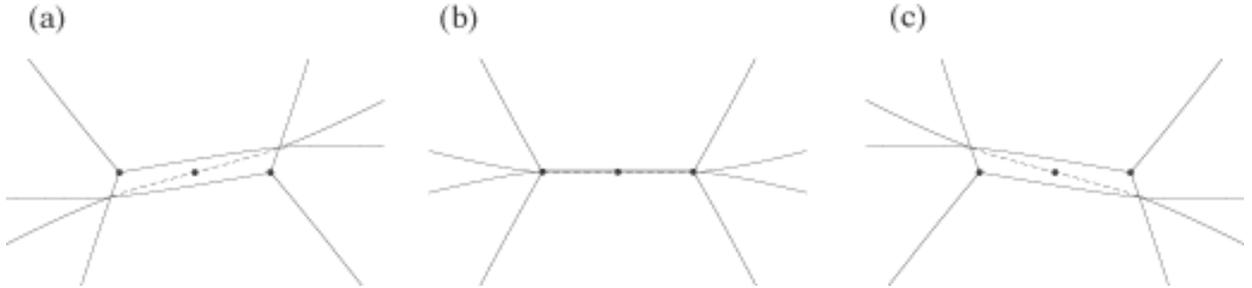}
\end{center}
\caption{
Stokes geometry of \eqref{eq:4} with
(a) $\textrm{arg} \eta = (\frac{1}{2} - \frac{1}{12})\pi$,
(b) $\textrm{arg} \eta = \pi/2$ 
 and 
(c) $\textrm{arg} \eta = (\frac{1}{2} + \frac{1}{12})\pi$.
}
\label{bnr2}
\end{figure}
%%%%%%%%%%%%%%%%%%%%%%%%%%%%%%%%%%%%%%%%%%%%%%%%%%%%%%%%%%%%%%%%%%%%%%%%%%%%%%%

By changing $\textrm{arg} \eta$ in Eq.\eqref{eq:4},
we find the Stokes geometry for 
(a) $\textrm{arg} \eta = (\frac{1}{2} - \frac{1}{12})\pi$,
(b) $\textrm{arg} \eta = \pi/2$  and
(c) $\textrm{arg} \eta = (\frac{1}{2} + \frac{1}{12})\pi$
respectively in Fig.~2 (a), (b) and (c).
Note that the virtual turning point is invariant when 
$\textrm{arg} \eta$ is changed,
just like an ordinary turning point is so.
One then observes in Fig.~2 (a) and (c) the interchange of the relative 
location of a Stokes curve emanating from an ordinary turning point
and that emanating from a virtual turning point.
One also finds that the bifurcation of 
a Stokes curve in Fig.~2 (b) should look awkward
if the Stokes curves emanating from the virtual turning point
were not included in Fig.~2 (a), (c).
Since the bifurcation of the Stokes curve in Fig.~2 (b)
is a consequence of the (square-root type) singularity of
$\xi_2(x)$ at $x = \pm 1$,
bifurcation of a Stokes curve of this kind is
a rather universal phenomenon.
Actually when a Stokes curve hits a simple turning point
with a change of a parameter, the Stokes curve bifurcates
as in Fig. 3(b).
Fig. 3(a) and (c) respectively show the configuration of
two simple turning points $s_1,s_2$, a virtual turning point $v$
and Stokes curves emanating from them before and after the hitting.
See \cite{7a} for the concrete computation in the example
of the Stokes geometry for the quantized H\'enon map.

%%%%%%%%%%%%%%%%%%%%%%%%%%%%%%%%%%%%%%%%%%%%%%%%%%%%%%%%%%%%%%%%%%%%%%%%%%%%%%%
\begin{figure}[h]
\begin{center}
\includegraphics[width=1.0\linewidth]{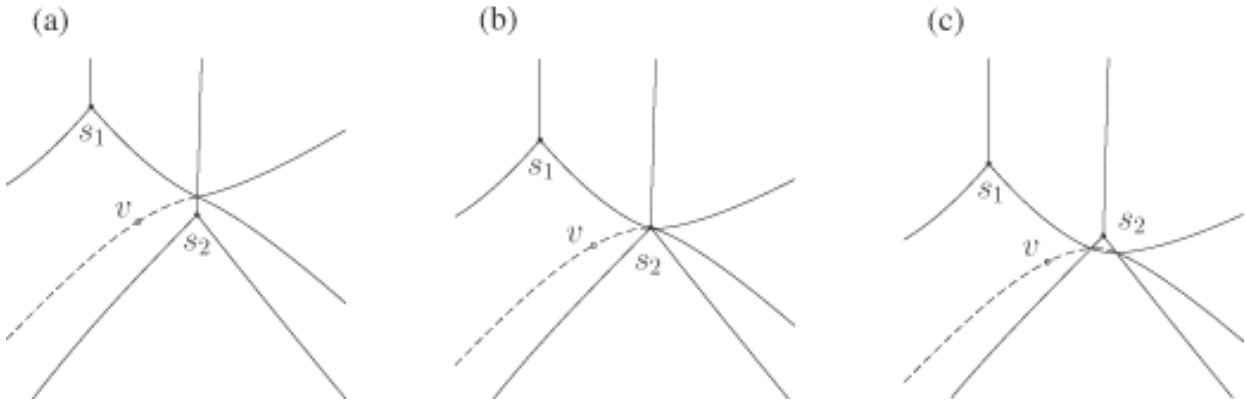}
\end{center}
\caption{
Schematic picture of bifurcation of the Stokes curve.
}
\label{bnr2}
\end{figure}
%%%%%%%%%%%%%%%%%%%%%%%%%%%%%%%%%%%%%%%%%%%%%%%%%%%%%%%%%%%%%%%%%%%%%%%%%%%%%%%

The relevance of virtual turning points and bifurcation
of Stokes curve so far described has turned out to bear
substantially important practical meaning 
in the exact WKB analysis (i.e., WKB analysis based
on the Borel resummation)
of the Painlev\'e hierarchy of 
Noumi-Yamada type \cite{5};
its first member consists of the symmetric form of 
the fourth Painlev\'e equation \cite{6}
\begin{align}\label{eq:7}
& \eta^{-1} \frac{df_{j}}{dt}=
f_{j}(f_{j+1}-f_{j+2})+\alpha_{j}
\quad (j=0,1,2) \\
\mbox{with} \ &
f_{j}=f_{j-3} \quad (j=3,4) \quad
\mbox{and} \quad  
\alpha_{0}+\alpha_{1}+\alpha_{2}=\eta^{-1},
\nonumber
\end{align}
and its underlying ``Schr\"odinger'' equation
\begin{equation}\label{eq:8}
-\eta^{-1}x \frac{\partial}{\partial x}
\begin{pmatrix}
\psi_{0}\\
\psi_{1}\\
\psi_{2}
\end{pmatrix}
=
\begin{pmatrix}
(2\alpha_{1}+\alpha_{2})/3 & f_{1} & 1\\
x  & (-\alpha_{1}+\alpha_{2})/3 & f_{2} \\
xf_{0} & x & -(\alpha_{1}+2\alpha_{2})/3
\end{pmatrix}
\begin{pmatrix}
\psi_{0}\\
\psi_{1}\\
\psi_{2}
\end{pmatrix}
\end{equation}
together with its appropriate 
deformation equation omitted here.

%%%%%%%%%%%%%%%%%%%%%%%%%%%%%%%%%%%%%%%%%%%%%%%%%%%%%%%%%%%%%%%%%%%%%%%%%%%%%%%
\begin{figure}[h]
\begin{center}
\includegraphics[width=1.0\linewidth]{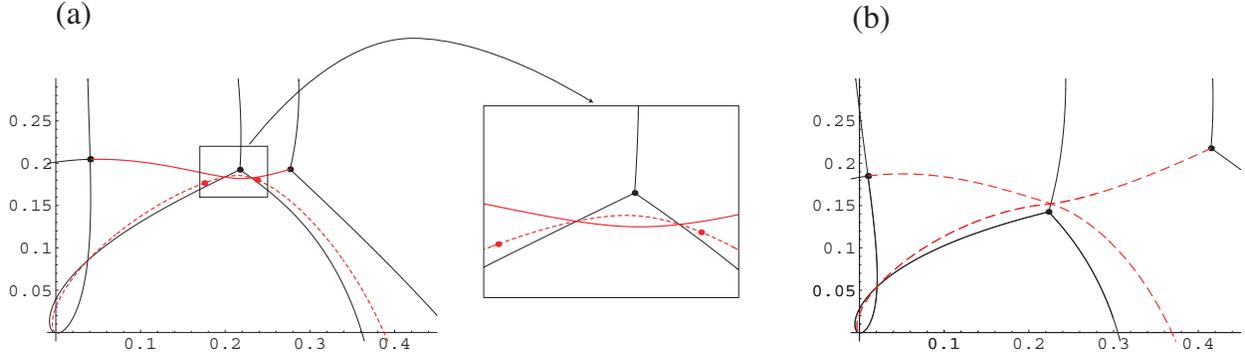}
\end{center}
\caption{
Fig. 4(a) shows the relevant part of 
the Stokes geometry of \eqref{eq:8}
when $t$ lies on a portion of 
a Stokes curve that contains its starting turning point
and Fig. 4(b) shows  that for $t$ 
lying on the Stokes curve but far away
from the turning point.
For the clarity of the presentation we use the following convention:
A black (resp. red) dot indicates an ordinary (resp. a virtual)
turning point.
A black (resp. red) solid line indicates a Stokes curve emanating
from an ordinary turning points (resp. doubly overlapped Stokes curves
from ordinary turning points) and a red dotted (resp. dashed) line
indicates a Stokes curve from a virtual turning point sitting
on a Stokes curve from another virtual (resp. ordinary) 
turning point.
}
\label{bnr2}
\end{figure}
%%%%%%%%%%%%%%%%%%%%%%%%%%%%%%%%%%%%%%%%%%%%%%%%%%%%%%%%%%%%%%%%%%%%%%%%%%%%%%%

A computer-assisted study \cite{7}
of the Stokes geometry of \eqref{eq:8}
shows that its expected degeneracy, i.e.,
the appearance of a Stokes curve
connecting two turning points
is not observed when the parameter $t$
enters on a some portion
of a Stokes curve of \eqref{eq:7}
if we take into account only ordinary turning points;
this contradicts with the result \cite{8}
obtained for the same fourth
Painlev\'e equation but with a different underlying
Schr\"odinger equation of the second order.
The paradoxical situation is resolved
if both ordinary and virtual turning points
are taken into consideration.
To illustrate the situation
we show the following Fig. 4;
In Figure 4(b) the ordinary simple turning point is connected with
a virtual turning point by a Stokes curve and 
the ordinary double turning point is connected with another
virtual turning point.

The point is that a Stokes curve hits a turning point
in the Stokes geometry of \eqref{eq:8} at some point $t=t_*$
on the Stokes curve of \eqref{eq:7} and that 
the role of an ordinary turning point and that of
a virtual turning point are switched there
through the mechanism we found in Fig. 3.

More complicated topological changes of
configurations of Stokes curves and turning points,
both ordinary and virtual, will be reported elsewhere
\cite{7a}, \cite{7};
they are basically attained by repeated applications of the mechanism 
observed in Fig.~3.

\begin{acknowledgements}
The research of the authors has been supported in part by JSPS
Grant-in-Aid No.~1434042, No.~15540190 and No.~14077213.
\end{acknowledgements}

\end{document}